\documentstyle[12pt,twoside,fleqn,espcrc1,epsf]{article}

\newcommand{\AmS}{{\protect\the\textfont2
  A\kern-.1667em\lower.5ex\hbox{M}\kern-.125emS}}

\title{Instanton-Induced Interactions in Finite Density QCD
}

\author{G. W. Carter\address{The Niels Bohr Institute, Blegdamsvej 17,
        DK-2100 Copenhagen, Denmark}
        and 
        D. Diakonov\address{NORDITA, Blegdamsvej 17,
        DK-2100 Copenhagen, Denmark}}

\begin{document}
\maketitle

\begin{abstract}
We consider the finite density, zero-temperature behaviour of quark matter
in the instanton picture. Since the instanton-induced interactions are
attractive in both $\bar{q}q$ and $qq$ channels, a competition ensues
between phases of matter with condensation in either or both.
It results in chiral symmetry restoration due to the onset of diquark
condensation, a `colour superconductor', at finite density.  
\end{abstract}

\section{Introduction} 

Due to a lack of lattice QCD techniques for implementing quark chemical 
potential, the finite density properties of strongly-interacting matter remain
unresolved.
To date, model studies suggest not only chiral symmetry restoration but also
the possibility of Cooper pairing of quarks
at high density, via an attractive $qq$ interaction, similar to 
superconducting electrons. 
The analogy has been extended to nomenclature,
with the QCD version called colour superconductivity.

It has been known for some years that perturbative, single
gluon exchange between quarks is attractive and will generate a 
pairing gap around the Fermi surface \cite{BL}.
More recently, it was suggested that colour superconductivity might also arise
by non-perturbative means at moderate quark density \cite{first}.  
Since then, more
detailed studies using models inspired by that of Nambu and Jona-Lasinio
\cite{BR} and instantons \cite{CD} have supported this idea.  This talk
describes how diquark formation restores broken chiral symmetry in the
context of the QCD instanton vacuum, an approach which has accounted for
many hadronic observables through the use of fundamental degrees of 
freedom (quarks and gluons) in a microscopic approximation.


\section{Quark Effective Action}

The derivation of an effective action for chiral quarks in $N_f$ flavours
has been discussed in detail in other publications.  Here we concentrate
on the two flavour case, which is often adequate for low energy phenomenology.
Growing quark chemical potential naturally makes the strange quark more
relevant, as has been studied in other models \cite{strange}.
These authors conclude that the two-flavour superconducting state is likely
to be present at moderate values of the quark chemical potential for a
realistic strange mass.

The basic idea is to replace the partition function of QCD with an effective
form which divides the low and high energy contributions.  The high
momentum part is taken to be perturbative and as such the gluons here are
assumed to be
small corrections to the stable, low-energy configurations of the gauge
fields -- the instantons.  Each (anti-) instanton in turn induces a quark zero
mode of (right) left chirality, and averaging over all possible instanton
backgrounds results in a delocalization of the zero modes which spontaneously
breaks chiral symmetry.

This picture of the vacuum is supported by
various lattice studies and has a long history of successful phenomenology.
Following this procedure, the expected 't Hooft interaction is obtained
and one has an effective quark action which is suitable for practical
calculations.  
We have reformulated this effective action for finite quark chemical potential. 
The result can be expressed as
\begin{eqnarray}
S[\psi,\psi^\dagger] &&\hskip-12pt = -\int\!d^4p\;\psi^{\dagger}( p\hskip-6pt/
+ i\mu\gamma_4) \psi + \lambda \Big(
 \int\!dU\;
\int\!\prod_f^{N_f}\left[\frac{(d^4p_fd^4k_f)}{(2\pi)^{8}}\right]
(2\pi)^4\delta^4\Big(\sum(p_f-k_f)\Big)
\nonumber\\
&&\hskip-1.6cm \prod^{N_f}_f \Big[
\psi^\dagger_{Lf\alpha_f i_f}(p_f)
{\cal F}(p_f,\mu)_{k_f}^{i_f}\epsilon^{k_fl_f}
U_{l_f}^{\alpha_f} U_{\beta_f}^{\dagger o_f} \epsilon_{n_fo_f}
{\cal F}^\dagger(k_f,-\mu)_{p_f}^{n_f}\psi_L^{f\beta_fp_f}(k_f)\Big]\,
 + {\rm (L \leftrightarrow R) }\Big)\,,
\label{vertex}
\end{eqnarray}
where the $\psi$ are quark fields and the ${\cal F}$ are spin/colour matrix
form factors.  They involve the quark zero modes and as such have specific
dependence on momentum and chemical potential \cite{CD}.
Unique to this approach, the coupling constant $\lambda$ is not fixed here.
One rather integrates over all possible coupling strengths since the 
constant has been introduced as a Lagrange multiplier.
This gives rise to a subtle interdependence
between the instanton background and the quark interaction.

\section{Results for Finite Quark Density}


With this effective theory it is straightforward to 
perturbatively expand in the coupling constant $\lambda$.
This amounts to a virial expansion in the instanton density $N/V$, 
with which $\lambda$ scales nonlinearly.
We consider only the scalar condensates, which translates into an ansatz
of three propagators: two normal and one Gorkov.
It allows condensation in the $\bar{3}$ diquark channel and breaks 
the colour symmetry as $SU(3)\rightarrow SU(2)\times U(1)$.
Should this occur, the normal propagators (and ensuing condensates) will lose
their colour degeneracy and 
their separation
becomes necessary.  Namely, there will be two gapped and one ungapped
quark species.  These propagators allow for chiral symmetry
breaking, as explained in Ref. \cite{CD}.

\begin{figure}[bt]
\setlength\epsfxsize{14cm}
\centerline{\epsfbox{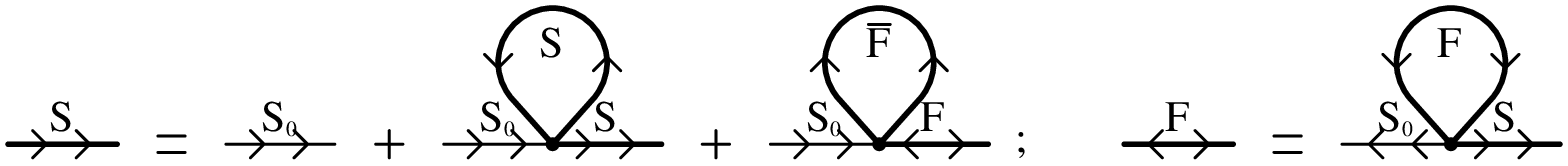}}
\vskip-5mm
\caption{Schwinger-Dyson-Gorkov diagrams to first order in $\lambda$.}
\label{gorkfig}
\vskip-1mm
\end{figure}
This procedure closes a set of Schwinger-Dyson-Gorkov equations, shown in Fig.
\ref{gorkfig}.  The diagrams reduce to three closed loops corresponding to
a set of three gap equations which
specify the condensates $g_1$, $g_2$, and $f$.  The first two are chiral
condensates, distinct in the case of colour symmetry breaking, and
the third a diquark condensate.  
%
They combine in the physical quantities:
$M_1 = \left(5-4/N_c\right)g_1+ \left(2 N_c-5+2/N_c\right)g_2$,
$M_2 = 2\left(2-1/N_c\right)g_1+2(N_c-2)g_2$, and
$\Delta = \left(1+1/N_c\right)f$.
The $M_{1,2}$ are measures of chiral symmetry breaking and
act as an effective mass.
Meanwhile the diquark loop $2\Delta$ plays the role of twice
the single-quark energy gap formed around the Fermi surface.

The solution of the gap equations depends on the vertex coupling constant,
$\lambda$, which itself is determined by balancing the instanton background
with the condensates through its saddle-point value.  This minimization of
the partition function leads to \cite{CD}
\begin{equation}
N/V = \lambda\langle Y^+ + Y^-\rangle = 4(N_c^2-1)
\left[ 2 g_1 M_1 + (N_c-2) g_2M_2 + 4 f\Delta \right]/\lambda\,.
\label{gapeqn}
\end{equation}
This joins the gap equations to close a system of equations, numerically
solvable.
Once this is done, the chiral condensate proper may be computed as an
integral over the resummed propagator.


For any given chemical potential, multiple solutions can be obtained for
the gaps.  These correspond to different phases of quark matter, and they
are summarized as follows:
(0)  Free massless quarks: $g_1=g_2=f=0$;
(1)  Pure chiral symmetry breaking: $g_1=g_2\ne 0$, $f=0$;
(2)  Pure diquark condensation: $g_1=g_2=0$, $f\ne 0$;
and (3)  Mixed symmetry breaking:  $g_1 \ne g_2 \ne 0$, $f\ne 0$.
The free energy, calculated to first order in $\lambda$, is minimized
in order to resolve the stable solution.
The phase corresponding to the {\em lowest} coupling $\lambda$
is the thermodynamically favoured \cite{CD}.

\begin{figure}[b]
\begin{minipage}[t]{80mm}
\setlength\epsfxsize{7.9cm}
\epsfbox{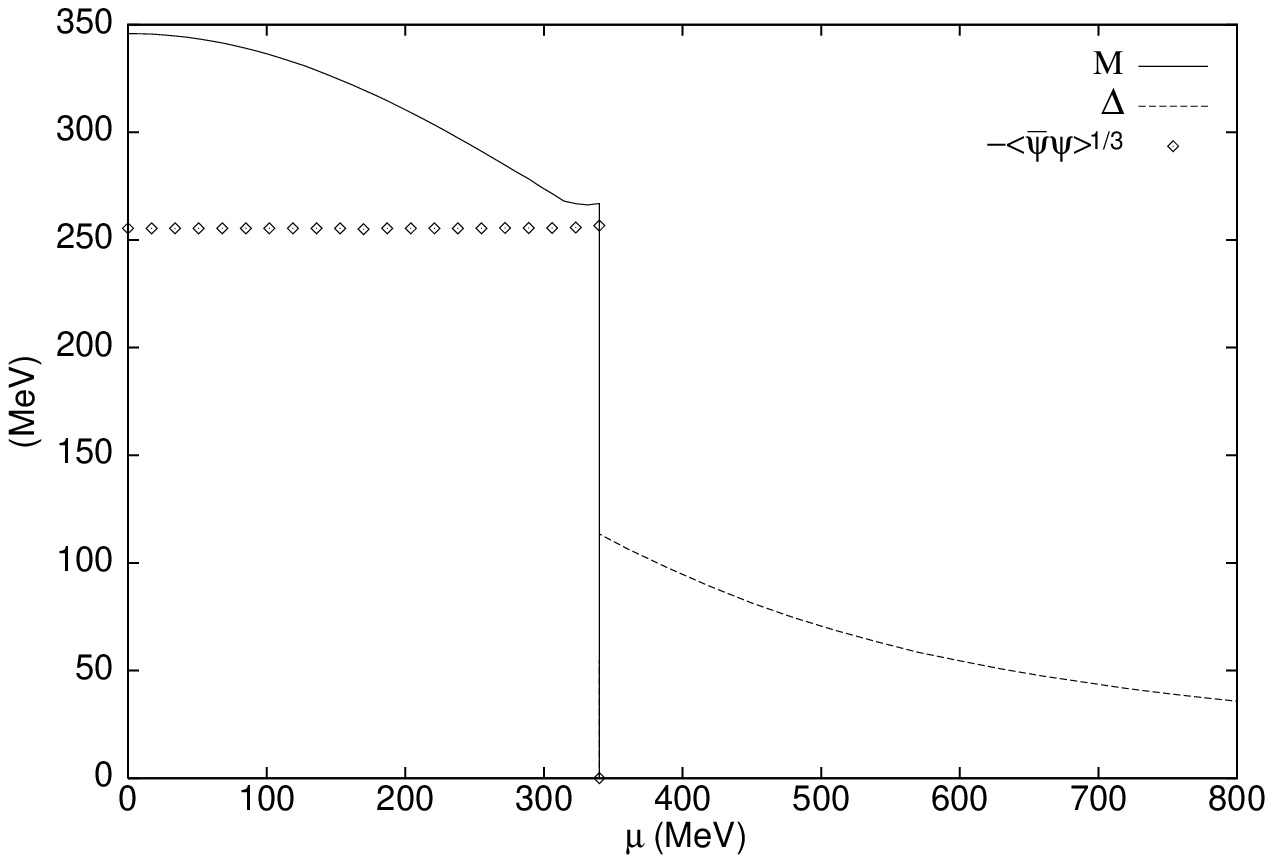}
\vskip-1cm
\caption{ Condensates for $N_c=3$ as a function of $\mu$.
}
\label{phasedia}
\end{minipage}
\hspace{\fill}
\begin{minipage}[t]{75mm}
\setlength\epsfxsize{7.4cm}
\epsfbox{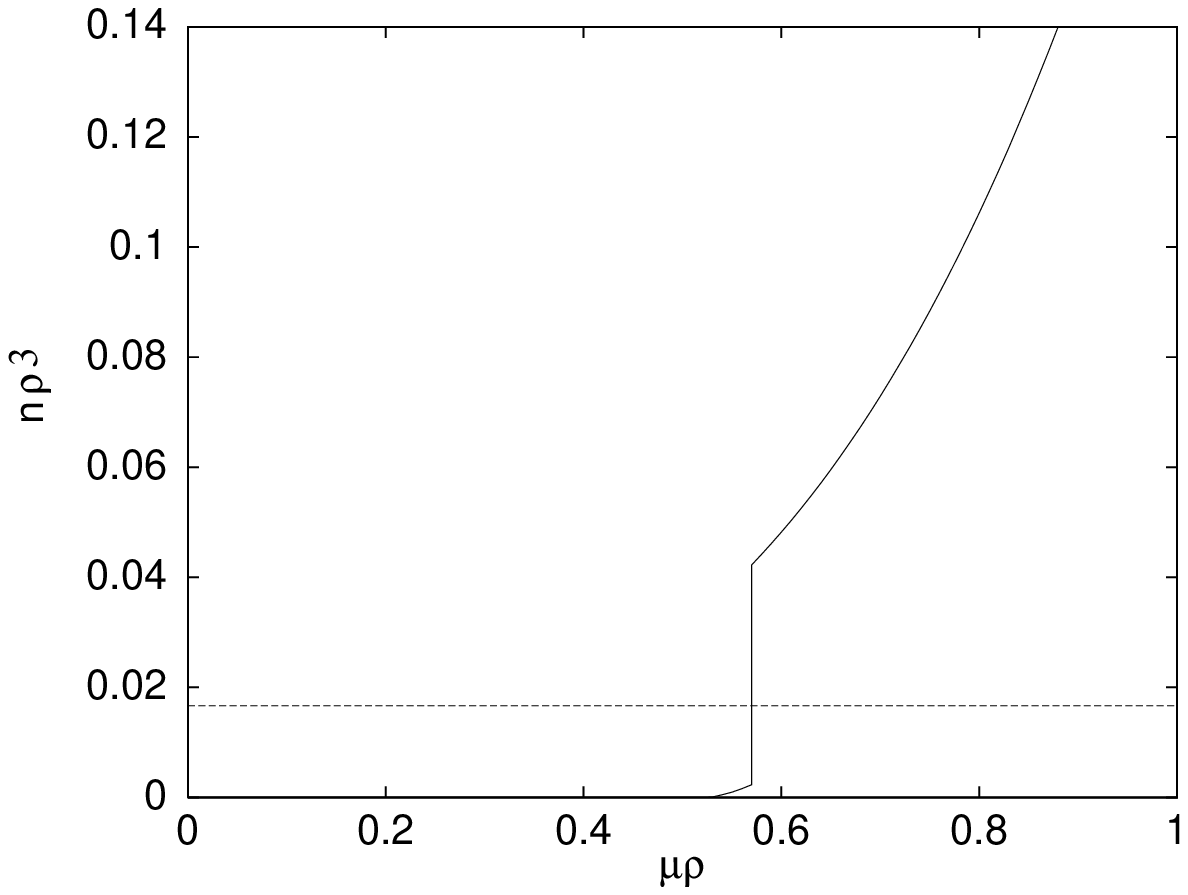}
\vskip-1cm
\caption{ The quark density $n_q$ vs. $\mu$.  }
\label{density}
\end{minipage}
\end{figure}

%
No solutions were found matching Phase (0), and the Phase (3) solution
obtained disappears at relatively low chemical potential ($\mu \approx
80$ MeV) and is never thermodynamically competitive \cite{CD}.  The remaining
phase competition is then between Phases (1) and (2).  In the vacuum, where
$\mu =0$, one finds Phase (1) preferred -- this is the standard picture.
However, at a critical chemical potential $\mu_c$, defined by the ratio
of superconductive gap to chiral effective masses
$\Delta/M = \sqrt{N_c/8(N_c-1)} = \sqrt{3}/4$,
a first-order phase transition occurs.  With the standard instanton parameters
$N/V = 1$ fm$^{-4}$ and $\bar\rho=0.33$ fm, we find $\mu_c\simeq 340$ MeV.
The first-order nature of the phase transition is clearly seen
in Fig. \ref{phasedia}.

\begin{figure}[bt]
\begin{minipage}[t]{80mm}
\setlength\epsfxsize{6.5cm}
\epsfbox{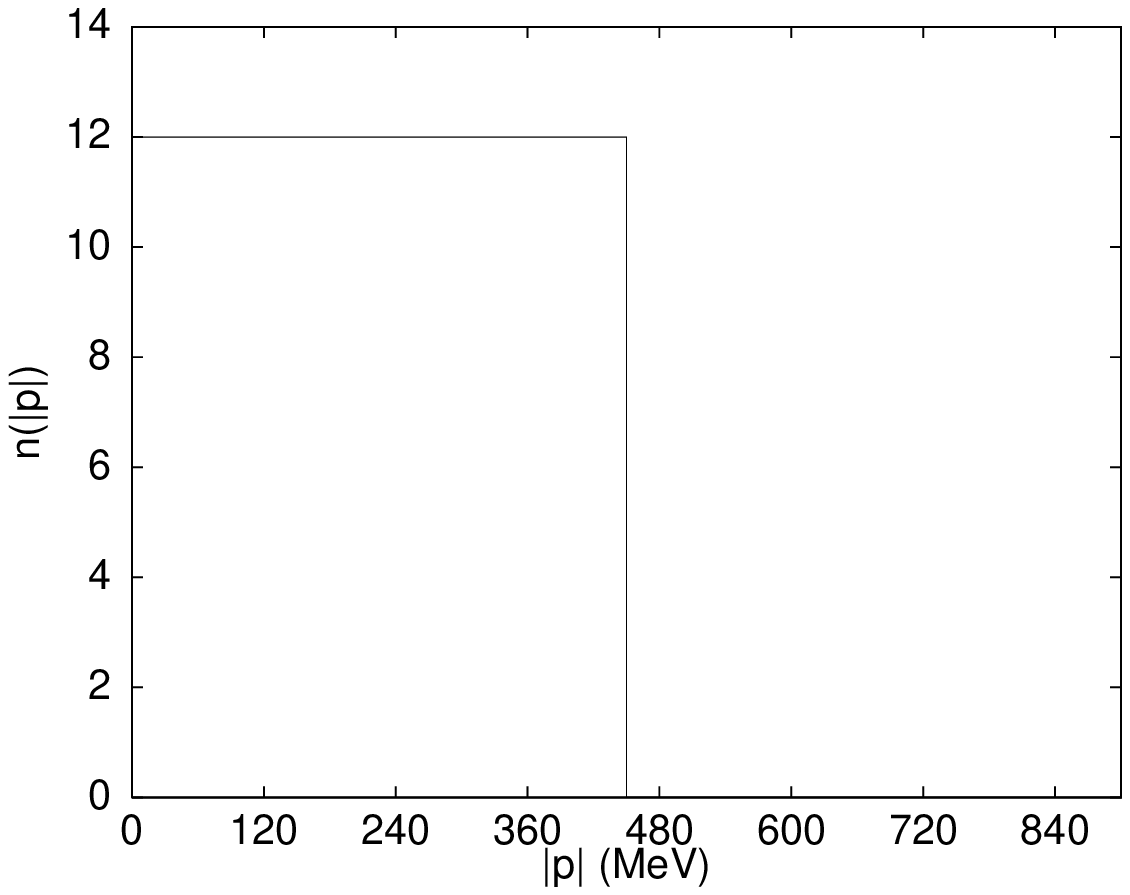}
\vskip-1cm
\caption{ 
Occupation number $n(p)$ vs. $p$ for Phase (1) for $\mu=1/\bar\rho=600$ MeV.}
\label{occ1}
\end{minipage}
\hspace{\fill}
\begin{minipage}[t]{75mm}
\setlength\epsfxsize{6.5cm}
\epsfbox{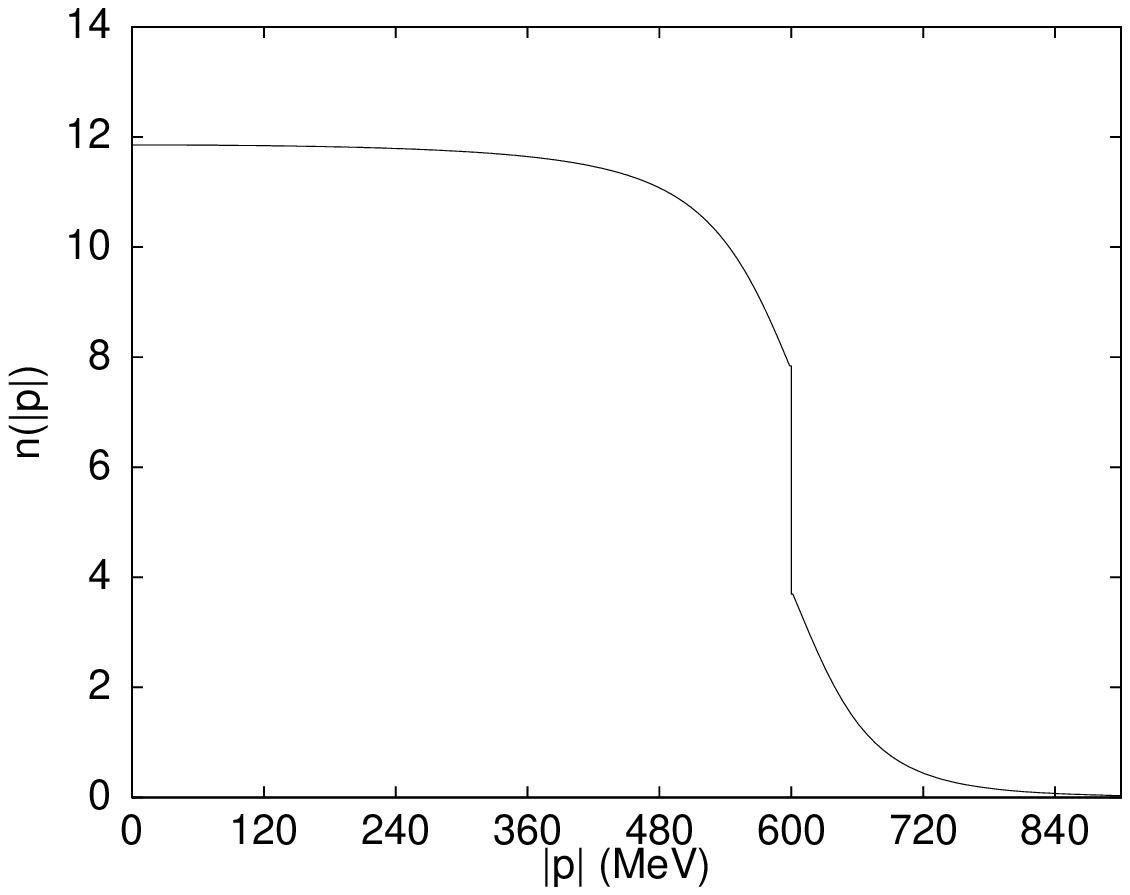}
\vskip-1cm
\caption{ 
Occupation number $n(p)$ vs. $p$ for Phase (2) for $\mu=1/\bar\rho=600$ MeV.}
\label{occ2}
\end{minipage}
\vskip-1mm
\end{figure}

Physically, the quark density is more relevant than the chemical potential.
As an intermediate step and in order to demonstrate the microscopic 
differences between the two phases, we have calculated the occupation number 
density for quarks.   This is nontrivial and here we present
only numerical results in Figs. \ref{occ1} and \ref{occ2}.
In Phase (1), there is clearly
an effective mass brought about by spontaneous symmetry breaking, indicated
by the reduced Fermi radius.  We stress that, despite the complicated
four-momentum dependence of the interaction, the
resulting occupation number density appears as a perfect Fermi step
function.  Cooper pairing, however, smears the Fermi
surface, and this is evinced in the second plot.  The residual discontinuity
at $|\vec{p}|=\mu$ is the contribution from the free, colour-3 quarks
which do not participate in the diquark.  

Integrating over momenta and recalling
the critical chemical potential, the quark density profile as a function
of chemical potential is plotted in Fig. \ref{density}
for the equilibrium states.  We see a 
discontinuity at the phase transition, where the horizontal line signifies
the quark density of stable nuclear matter.  The phase transition
occurs at an extremely low quark density, which remains a conceptual 
conundrum.

\section{Conclusions}

Beginning from the instanton picture of the QCD vacuum, we have extended the
model for finite density and found chiral symmetry restoration due to 
the onset of colour superconductivity.  This phase transition is strongly
first order and in agreement with other quark-based approaches.

\end{document}